\documentclass[prb,twocolumn]{revtex4}
\newcommand{\bea}{\begin{eqnarray}}
\newcommand{\eea}{\end{eqnarray}}
\newcommand{\ben}{\begin{equation}}
\newcommand{\een}{\end{equation}}
\newcommand{\benu}{\begin{enumerate}}
\newcommand{\enu}{\end{enumerate}}
\newcommand{\la}{\langle}
\newcommand{\ra}{\rangle}

\newcommand{\al}{\alpha}
\newcommand{\be}{\beta}
\newcommand{\dl}{\delta}
\newcommand{\ga}{\gamma}
\newcommand{\si}{\sigma}
\newcommand{\ld}{\lambda}
\newcommand{\ka}{\kappa}
\newcommand{\tht}{\theta}
\newcommand{\om}{\omega}
\newcommand{\ep}{\epsilon}
\newcommand{\ham}{\mathcal{H}}
\newcommand{\ord}{\mathcal{O}}
\newcommand{\prm}{\prime}

\newcommand{\br}{{\bf r}}
\newcommand{\bR}{{\bf R}}
\newcommand{\cda}{c^{\dagger}}
\newcommand{\num}{\mathcal{N}_s}
\newcommand{\cst}{\mathcal{C}}

\begin{document}

\title{Construction of Localized Basis for Dynamical Mean Field Theory}
\date{\today}

\author{I. Paul$^1$, and
G. Kotliar$^2$
}                   

\affiliation{
$^1$SPhT, CEA-Saclay, L'Orme des Merisiers, 91191 Gif-sur-Yvette, France.\\
$^2$Center for Materials Theory, Department of Physics and Astronomy,\\
Rutgers University, Piscataway, New Jersey 08854.
}

%
\begin{abstract}
Many-body Hamiltonians obtained from first principles 
generally include all possible 
non-local interactions. But in dynamical mean field theory the 
non-local interactions are ignored,
and only the effects of the local interactions are taken into account. 
The truncation of the non-local interactions is a basis dependent 
approximation.
We propose a criterion to construct an appropriate 
localized basis in which the truncation can be carried out. 
This involves finding a
basis in which a functional given by the sum of the squares of 
the local 
interactions with appropriate weight factors is maximized under unitary
transformations of basis. 
We argue that such a 
localized basis is suitable for the application of dynamical mean field theory
for calculating material properties from first principles.
We propose an algorithm which can be used for constructing the 
localized basis. We test our criterion on a toy model and find it 
satisfactory. 
\end{abstract}
%
\maketitle

\section{Introduction}
\label{sec:intro}
In the last decade and a half dynamical mean field theory (DMFT) has emerged
as an important tool for studying condensed matter systems with strong 
correlation.~\cite{gabi} 
The principal difficulty in understanding these systems is the 
non-perturbative character of such systems,
for which the physical properties 
cannot be understood by expanding various quantities in powers of the 
interaction. In this respect DMFT is a powerful tool for studying
problems of interacting electrons on a lattice. It is a non-perturbative 
technique which is able to capture fully the local  
dynamical correlations in the system. 
Single site DMFT, as an approximation scheme, is controlled in that the 
result is exact in the limit of large coordination numbers.~\cite{gabi}
Recent extensions of DMFT to clusters seem to be rapidly convergent 
for local observables.~\cite{biroli} 
Other cluster schemes such as cluster perturbation theory and dynamical
cluster approximation are also being used to study problems of strong electron
correlation.~\cite{dca-rmp}
Recently it
has also been recognized that DMFT can be used as a powerful  tool for the
realistic computation of properties of materials as in the
LDA+DMFT scheme.~\cite{anisimov,held,lichtenstein}
Indeed results for a large variety of materials ranging
from Cerium,~\cite{held2} Iron and Nickel,~\cite{imseok} 
Plutonium~\cite{nature} and many other oxides have
been successfully studied with this method  starting from first principles. 

A common way to utilize DMFT in first principles calculations is to
first derive a Hamiltonian with a kinetic energy part and a general
short-range interaction part. This Hamiltonian, which will be the starting point of this
paper in Eq.~(\ref{eq:ham1}), is subsequently studied  
by DMFT.
The long-range part of the Coulomb interaction can be taken into account by 
several means. For example, in extended DMFT this is done by coupling the 
electron at the impurity site to a bath of bosons whose spectral function is 
determined self-consistently.~\cite{si,chitra} 
This is equivalent to treating the electrons in the presence
of a fluctuating electric field (long range interaction). Another possibility is to follow
along the lines of Bohm and Pines.~\cite{bohm-pines}
In this method, starting with the charged electron gas, one performs canonical 
transformations to screen the electrons. In the resulting Hamiltonian the 
excitations are no longer the bare charged electrons, but screened neutral 
quasi-particles. 
There are various methods to obtain the starting Hamiltonian for DMFT.
(1) In one of the approaches the kinetic
energy term is the Kohn Sham Hamiltonian of a density functional theory
calculation written in a local basis set. The interaction terms, which can
include on-site (Hubbard ) as well as the short range part of the Coulomb
interaction, is evaluated using constrained LDA.~\cite{anisimov}
(2) In an alternative procedure, as mentioned above, 
one could start with the electron gas
Hamiltonian and the periodic potential, and 
perform the Bohm-Pines canonical
transformation~\cite{bohm-pines} to reduce the range of the Coulomb 
interactions, and 
then write
the transformed Hamiltonian in a local basis set. (3) A third approach
proposed recently~\cite{ferdi} uses the GW approach to obtain the interaction
strength.

The next step is the study of the resulting Hamiltonian with a short-range
interaction using DMFT.
This involves
local approximations, and the notion of locality
depends explicitly  on the basis set considered.
To illustrate the point, if we 
perform an invertible transformation of
the original basis, we merely re-express the original Hamiltonian in a new
basis, provided we keep all the terms in the Hamiltonian. The full 
electron Green's
function is obtained by applying the same transformation to the creation
and destruction operators.
 But in practice, 
one  performs two approximations that explicitly depend on the
basis set. The first one is to neglect interactions whose range exceeds the
cluster size (truncation).  
The second (local approximation) consist of setting equal to
zero the elements of the self energy which exceeds that size.
These two approximations explicitly depend on the definition of locality
which is encoded in the basis set. In this paper we address only
the first issue,
and argue that 
truncating non-local 
interactions is appropriate when the wave-functions of the basis are 
well localized. 
As DMFT techniques are beginning  to be applied to Hamiltonians with
realistic interactions involving non-local terms,~\cite{schiller}
there is need for 
well-defined criteria for choosing optimal bases for 
computations.
The purpose of this paper is to propose one
criterion which can be used to construct a localized basis for
DMFT computations. 

The method of choosing a suitable localized basis of wave-functions has been 
studied 
earlier in quantum chemistry and in band structure theory.~\cite{marzari}
The formulation of
the problem consists of two steps. First, one identifies a certain group of 
transformations of the basis states, say for example, unitary
transformations. Second, one identifies a criterion that picks out one basis
out of all possible choices that are connected by the transformations. The 
criterion is a basis dependent quantity, and therefore is a functional in the 
space of the transformations. It is a measure of the amount of 
localization of the wave-functions in a given basis. For example, in quantum
chemistry ``energy localized molecular orbitals'' have been 
studied.~\cite{edmiston} 
These
are obtained by maximizing under unitary transformations a functional 
given by 
the sum of the Coulomb self-interaction of the orbitals. Similarly,
for band structure calculations the use of ``maximally-localized'' Wannier
functions has been proposed.~\cite{marzari} 
The idea is to exploit the freedom that is 
present in the choice of the phases of the Bloch orbitals. 
With a given set of Bloch orbitals one can define a new set by a unitary 
transformation. From each such set of Bloch orbitals one can obtain a 
corresponding set of Wannier functions by Fourier transformation. The 
maximally-localized Wannier functions are obtained by minimizing the spread
functional, which is the sum of the second moments of the Wannier functions,
in the space of unitary transformations. More recently, the construction of
localized basis states has been extended to include non-orthogonal 
molecular orbitals.~\cite{liu}


The rest of the paper is organized as follows. In section \ref{sec:basis} we 
identify a criterion for choosing a basis suitable for DMFT. We construct a 
functional which is maximum in the preferred basis. We discuss the
properties of such a basis by studying linear variations of the functional 
under unitary transformations. We also propose a method for 
constructing the preferred basis. In section \ref{sec:conclusion} we test the
criterion on a Hamiltonian whose interaction is taken to be simple but non-trivial.
 We find that the criterion and the associated 
functional is well-behaved. In conclusion, we summarize our 
main results. 

\section{Localized basis for DMFT}
\label{sec:basis}

To keep the discussion general, in the following we formulate the problem
in a basis which is non-orthogonal. For this purpose we  consider a 
system of interacting electrons on a lattice whose Hamiltonian is expressed in
a basis of atomic orbitals. The single particle states are denoted by
$\phi_{\al}(\br - \bR_n) \equiv \la \br | n \al \ra$, where $\al$ is a symmetry
related index (say, orbital) and $\bR_n$ is a lattice position. We 
suppose there are $m$ orbitals per site such that the 
index $\al = 1, \cdots , m$, and 
there are $N$ lattice sites with the index $n = 0, \cdots , N-1$. We also
impose periodic boundary condition $| n, \al \ra = | n+N, \al \ra$. The states
defining the basis, unlike those in a Wannier basis, are not orthogonal. We 
denote the overlap between any two states by $O_{\al \be}(n-m)
\equiv \la n \al | m \be \ra$. The second quantized many-body Hamiltonian can
be written as
\ben
\label{eq:ham1}
\ham = \sum_{n m \atop \al \be} t^{nm}_{\al \be} \cda_{n, \al} c_{m, \be}
+ \sum_{nmlk \atop \al \be \ga \dl} V^{nmkl}_{\al \be \dl \ga} \cda_{n, \al}
\cda_{m, \be} c_{k, \dl} c_{l, \ga}.
\een
We assume that the matrix elements 
$t^{nm}_{\al \be} \equiv \la n \al | \ham_0 | m \be \ra$ for the
non-interacting part, and 
$V^{nmkl}_{\al \be \dl \ga} \equiv \la n \al, m \be | \hat{V} | 
l \ga, k \dl \ra$ for the interacting part are known from first principles
studies such as band structure calculations. It is useful to bear in mind that
the anti-commutation relation between the creation and annihilation operators
in a non-orthogonal basis is given by 
$\{ \cda_{n, \al} \; , \; c_{m, \be} \} = O^{-1}_{\al \be}(n-m)$.~\cite{wegner}
We now consider an invertible transformation of the single particle basis
that preserves the lattice translation invariance,
$|n \al \ra \rightarrow |n^{\prm} \al^{\prm} \ra = \sum_{m \be}
T_{\be \al^{\prm}}(m-n^{\prm}) |m \be \ra$. Expressed in the new basis the 
Hamiltonian, say $\ham^{\prm}$, has the same form as in Eq.~(\ref{eq:ham1})
except with all the indices primed. We know that $\ham^{\prm} = \ham$, since
it is the same operator expressed in two different bases. However, 
when we
truncate all the non-local interactions, we deal with a model
Hamiltonian of the form
\ben
\label{eq:hamloc}
\ham_{{\rm tr}} = \sum_{n m \atop \al \be} t^{nm}_{\al \be} \cda_{n, \al} 
c_{m, \be}
+ \sum_{n \atop \al \be \ga \dl} V^{nnnn}_{\al \be \dl \ga} \cda_{n, \al}
\cda_{n, \be} c_{n, \dl} c_{n, \ga}.
\een
But the process of truncation is a basis dependent step. If we perform the 
truncation in the new basis, i.e., on $\ham^{\prm}$, the resulting new 
truncated Hamiltonian $\ham^{\prm}_{{\rm tr}} \neq \ham_{{\rm tr}}$. This 
observation implies that ignoring non-local interactions is a good
approximation only if the single particle basis is sufficiently localized. 
In the following we develop a systematic criterion for constructing such a 
basis.

Here we consider only unitary transformations of basis. Later we 
comment about the possibility of extending the scheme to include non-unitary
invertible transformations as well.
We start from an
initial basis $\{ | n \al \ra \}$, and consider unitary transformations
\ben
| n \al \ra \rightarrow | n^{\prm} \al^{\prm} \ra  = U | n \al \ra = 
\sum_{m \be} U_{\be \al}(m-n) | m \be \ra
\een
to new basis states $\{ | n^{\prm} \al^{\prm} \ra  \}$. In order to find a 
criterion to choose the most localized basis among the possible bases 
$\{ | n^{\prm} \al^{\prm} \ra  \}$, we first identify a quantity which is 
invariant under unitary transformations. The trace of any operator has this 
property. Since we are concerned about truncating the interacting part of the 
Hamiltonian, we consider the trace of the square of the interaction 
operator. In terms of the overlap matrix and the interactions expressed in the 
$\{ | n \al \ra \}$ basis this is given by 
\bea
I = {\rm Tr}(\hat{V}^2) 
&=& 
O^{-1}_{\al \be} (n-m) O^{-1}_{\ga \dl}(l-k)
O^{-1}_{\si \rho}(r-s) 
\nonumber \\
&\times&
O^{-1}_{\eta \nu}(p-q) 
V^{mkpr}_{\be \dl \eta \si}
V^{sqln}_{\rho \nu \ga \al}.
\eea
Here, and in the rest of the paper we adopt the convention that
repeated indices are summed. The invariant defined above has two basis
dependent parts, namely, terms that involve only the local interactions and 
those involving non-local interactions. Keeping only the local interactions in
a given basis, we define the ``local interaction functional''. For example, in
the basis $\{ | n \al \ra \}$ the functional has the value
\[
F[\{| n \al \ra \}] = O^{-1}_{\al \be} (0) O^{-1}_{\ga \dl}(0)
O^{-1}_{\si \rho}(0) O^{-1}_{\eta \nu}(0) V^{0000}_{\be \dl \eta \si}
V^{0000}_{\rho \nu \ga \al}.
\]
To elucidate the structure of the functional we first note 
that the overlap matrix remains unchanged under unitary transformations, i.e.,
\ben
\la n^{\prm} \al^{\prm} | m^{\prm} \be^{\prm} \ra = O_{\al^{\prm} \be^{\prm}}
(n^{\prm} - m^{\prm}) = \la n \al | m \be \ra = O_{\al \be}(n-m).
\een
Next, the transformation of the interaction terms is given by
\bea
V^{nmkl}_{\al \be \dl \ga} \rightarrow V^{n^{\prm} m^{\prm} k^{\prm} 
l^{\prm}}_{\al^{\prm} \be^{\prm} \dl^{\prm} \ga^{\prm}}
&=& 
U^{\ast}_{\si \al}(r-n) U^{\ast}_{\rho \be}(s-m) V^{rsqp}_{\si \rho
\nu \eta} 
\nonumber \\
&\times&
U_{\eta \ga}(p-l) U_{\nu \dl}(q-k).
\eea
In terms of the unitary transformations the local interaction functional can
be written as
\bea
\lefteqn{
F[\{| n^{\prm} \al^{\prm} \ra \}] 
}
\nonumber \\
&&
= O^{-1}_{\al^{\prm} \be^{\prm}} (0) O^{-1}_{\ga^{\prm} \dl^{\prm}}(0)
O^{-1}_{\si^{\prm} \rho^{\prm}}(0) O^{-1}_{\eta^{\prm} \nu^{\prm}}(0) 
V^{0000}_{\be^{\prm} \dl^{\prm} \eta^{\prm} \si^{\prm}}
V^{0000}_{\rho^{\prm} \nu^{\prm} \ga^{\prm} \al^{\prm}}
\nonumber \\
&& =
\left[ O^{-1}_{\al \be} (0) O^{-1}_{\ga \dl}(0)
O^{-1}_{\si \rho}(0) O^{-1}_{\eta \nu}(0) \right]
\nonumber \\
&& \times
\left[ U^{\ast}_{\mu \be}(r) U^{\ast}_{\ka \dl}(s) V^{rsqp}_{\mu \ka \ld \tau}
U_{\tau \si}(p) U_{\ld \eta}(q) \right]
\nonumber \\
&& \times
\left[ U^{\ast}_{\pi \rho}(n) U^{\ast}_{\phi \nu}(m) V^{nmkl}_{\pi \phi \om
\tht} U_{\tht \al}(l) U_{\om \ga}(k) \right].
\eea
The inverse of the overlap matrix enters as weight factor, and the
interaction terms in the starting basis $\{ | n \al \ra \}$ serve as 
parameters of the functional. The desired basis is the one in which the 
functional is maximum in the space of unitary transformations. This criterion 
also implies that, in the chosen basis, the part of the invariant $I$ that 
contains non-local interactions is minimized.

In order to study the property of the preferred basis we consider 
infinitesimal unitary transformation given by $U = e^{i \ep H}$, where 
$H$ is hermitian and $\ep$ is a small parameter. The action of $H$ on the 
single particle wave-functions is given by 
$H | n \al \ra = H_{\be \al}(m-n) | m \be \ra$, such that
\bea
U_{\al \be}(n-m) 
&=& 
\dl_{\al \be} \dl_{nm} + (i \ep) H_{\al \be}(n-m) 
\nonumber \\
&+&
\frac{(i \ep)^2}{2!} H_{\al \ga}(n-l) H_{\ga \be}(l-m) + \cdots.
\nonumber
\eea
The hermiticity of $H$ implies that
\[ 
[ \la n \al | H | m \be \ra ]^{\ast} = \la m \be | H | n \al \ra , 
\]
i.e.,
\[
H^{\ast}_{\ga \be}(l-m) O_{\ga \al}(l-n) =
O_{\be \ga}(m-l) H_{\ga \al}(l-n).
\]
For a lattice of $N$ sites with periodic boundary condition and $m$ orbitals
per site, we note that the transformation matrix $H$ has $Nm^2$ real 
independent parameters. 
In the following we assume that $\hat{V}(\br_1, \br_2) 
= \hat{V}(\br_2, \br_1)$, so that $V^{nmkl}_{\al \be \dl \ga}
= V^{mnlk}_{\be \al \ga \dl}$. For the convenience of notation we define the 
quantity
\ben
\label{eq:Lmatrix}
L_{\si \mu}(t) \equiv O^{-1}_{\si \rho}(0) O^{-1}_{\al \be}(0) 
O^{-1}_{\ga \dl}(0) O^{-1}_{\eta \nu}(0) V^{0000}_{\rho \nu \ga \al}
V^{000t}_{\be \dl \eta \mu}.
\een
To $\ord (\ep)$ the variation of the functional can be written as 
\bea
\dl F &=& (- 4 i \ep) \left[ L^{\ast}_{\si \mu}(t) H^{\ast}_{\mu \si}(t) 
- L_{\si \mu}(t)
H_{\mu \si}(t) \right] 
\nonumber \\
&=& 
(- 4 i \ep) \left[ O^{-1}_{\si \be}(m-n) L^{\ast}_{\al \be}
(N-m) O_{\al \mu}(n-t) 
\right.
\nonumber \\
&& \left.
- L_{\si \mu}(t)
\right] H_{\mu \si}(t).
\eea
We define
\ben
\label{eq:Amatrix}
A_{\si \mu}(t) \equiv L_{\si \mu}(N-t) - O^{-1}_{\si \be}(m-n) 
L^{\ast}_{\al \be}(N-m) O_{\al \mu}(n+t),
\een
and we note that $A$ is anti-hermitian, i.e.,
\ben
\label{eq:anti-h-A}
O_{\be \ga}(m-l) A_{\ga \al}(l-n) = - A^{\ast}_{\ga \be}(l-m) O_{\ga
\al}(l-n).
\een       
The condition for the functional $F$ to have a local maximum is
\ben
\frac{\dl F}{\dl H_{\mu \si}(t)} = A_{\si \mu}(N-t) = 0.
\een
The above anti-hermitian condition has to be satisfied by the preferred
basis. In other words, the preferred basis is the one in which 
$L_{\si \mu}(t)$ is hermitian. The above condition gives $Nm^2$ real
independent equations, which is the same as the number of real independent 
parameters in the transformation matrix $H$.

The following is a simple ansatz for maximizing $F$ by successive unitary 
transformations. We start with an initial basis $\{ | n \al \ra \}$, and we 
calculate $A_{\mu \si}(t)$ using Eqs. (\ref{eq:Lmatrix}) and
(\ref{eq:Amatrix}). We then change the basis using the transformation
\ben
H_{\mu \si}(t) = i A_{\mu \si}(t),
\een
and iterate this procedure until the condition for the maximum is achieved. We
assert (proved in the appendix) that with this ansatz, to $\ord(\ep)$
\ben
\label{eq:assert}
\dl F = - (4 \ep) A_{\si \mu}(N-t) A_{\mu \si} (t) \geq 0.
\een
This ensures that with successive transformations the value of the 
functional increases (provided $\ep$ is small enough) until it reaches a local
maximum.

Our method for constructing a localized basis is similar to Wegner's flow 
equation~\cite{wegner2}
and Glazek and Wilson's similarity renormalization scheme~\cite{wilson}
 approaches where
small unitary transformations are used to reduce non-diagonal elements of a 
Hamiltonian. In our case, we use unitary transformations to reduce the strength
of non-local interactions and increase the value of the functional $F$.

\section{Discussion and Conclusion}
\label{sec:conclusion}

We test our criterion on a lattice Hamiltonian with an interaction of 
the form
\ben
\ham_{\rm int} = \sum_{ijkl} U_{ijkl} \cda_i \cda_j c_k c_l,
\een
where
\ben
\label{eq:toy-int}
U_{ijkl} =\sum_{\bR} \phi^{\ast}(i -\bR) \phi^{\ast}(j - \bR)
\phi(k -\bR) \phi(l - \bR).
\een
Here $\phi(i -\bR) \equiv \phi(\bR_i -\bR)$ is an orbital on lattice site 
$\bR_i$, which defines a one-particle starting basis $\{ \phi(n - \bR) \}$.
The interaction described above is local in configuration space, but in the 
basis of $\{ \phi(n - \bR) \}$ it has non-local terms as well.
For simplicity we consider one orbital per site, and assume that the starting
basis is orthonormal. In this basis the local interaction functional is given
by
\ben
F = \frac{1}{\num} \sum_{n, \bR, \bR^{\prm}} \left| \phi(n -\bR) \right|^4
\left| \phi(n -\bR^{\prm}) \right|^4,
\een
where $\num$ is the number of lattice sites. We note that the above 
functional has a form similar to the (square of the) 
inverse participation ratio that is studied in the context of Anderson 
localization.~\cite{thouless}  
We perform a unitary transformation of the basis of the form
$\phi(n - \bR) \rightarrow \phi(n - \bR) + \dl \phi(n - \bR) $,
where
\[
\dl \phi(n - \bR) = (i \ep) \sum_m h_{mn} \phi (m -\bR) + \ord(\ep^2).
\]
The coefficients $h_{mn}$ are hermitian such that $h^{\ast}_{mn} = h_{nm}$.
The variation of the functional $F$ can be written as
\[
\dl F = \frac{4 \cst}{\num} \sum_{n, \bR} \left| \phi(n - \bR) \right|^2
\left\{ \phi^{\ast}(n - \bR) \dl \phi (n -\bR) + {\rm h.c.} \right\},
\]
where $\cst = \sum_{\bR} \left| \phi (n - \bR) \right|^4$ is a site 
independent constant. Using the above form for $\dl \phi$ and the unitarity
of the transformation, we get to $\ord(\ep)$
\bea
\dl F &=& 
\frac{4 i \ep \cst}{\num} \sum_{n, m, \bR} h_{nm} 
\phi^{\ast} (m - \bR) \phi (n - \bR) 
\nonumber \\
&\times&
\left[ \left| \phi (m - \bR) \right|^2
- \left| \phi (n - \bR) \right|^2 \right].
\eea
The extrema of the functional is given by
\ben
\sum_{\bR} \phi^{\ast} (m - \bR) \phi (n - \bR)
\left[ \left| \phi (m - \bR) \right|^2
- \left| \phi (n - \bR) \right|^2 \right] = 0,
\een
for all sites $(n, m)$. By inspection, there are two solutions to the 
above equation. (1) $ \left| \phi (n - \bR) \right|^2 = 1/\num, \forall n$,
which is the limit of delocalized states, for which $F = 1/\num^2$ 
(minimum). (2) $ \left| \phi (n - \bR) \right|^2 =  \dl_{n, \bR}$, which is 
the limit of localized states. In this case $F = 1$ (maximum), and the 
interaction is entirely on-site. Starting with the original basis 
$\{ \phi(n - \bR) \}$, and the ansatz
\bea
h_{nm} &=& i \sum_{\bR} \phi^{\ast} (n - \bR) \phi (m - \bR)
\nonumber \\
&\times&
\left[ \left| \phi (n - \bR) \right|^2
- \left| \phi (m - \bR) \right|^2 \right],
\eea
we get to $\ord(\ep)$
\bea
\dl F 
&=& 
\frac{4  \ep \cst}{\num} \sum_{n, m, \bR} 
\left| \phi^{\ast} (n - \bR) \phi (n - \bR) \right|^2
\nonumber \\
&\times&
\left[ \left| \phi (m - \bR) \right|^2
- \left| \phi (n - \bR) \right|^2 \right]^2
\nonumber \\
\geq 0.
\eea 
Provided $\ep$ is chosen small enough (to justify the neglect of 
higher order variations), with the above ansatz it is possible to 
increase the value of $F$ with successive unitary transformations until
the limit of localization is attained. This simple example illustrates 
how the local interaction functional can be used to construct a basis of 
localized one-particle states. For interactions which are more complicated and 
realistic, it is unlikely that unitary transformations can make the interactions 
entirely on-site. However, the strengths of the non-local terms can be reduced
(quantitatively defined by maximization of functional $F$) in a more localized 
basis.

Two more comments are of relevance. First, our criterion 
ignores the non-interacting part of the Hamiltonian. If one starts
with nearest neighbour hopping in the original basis, in
the localized basis the hopping will be more complicated. But the point
of view adopted here is that
the non-interacting part can still
be solved exactly. 
Second, in this paper we consider only unitary transformations of
basis. This implies that one maximizes the local interaction functional within
a family of bases with the same overlap matrix (say, orthonormal bases, if 
the original basis is orthonormal). In principle one could probe for bases
with different overlap matrices by general invertible 
transformations. Such a group is non-compact and one needs to 
impose constraints such that the functional is 
bounded from above. One possible constraint can be imposed in terms of the 
singular value decomposition of the transformation matrix, say, the ratio of
the maximum and the minimum singular values be within a specified bound. 

In conclusion, we propose a criterion for constructing 
a localized single particle basis where non-local interactions
can be truncated.
Such a basis is 
appropriate for using DMFT for the calculation of material 
properties. We suggest a simple algorithm by which the construction
of the localized basis can be carried out. By testing  
the criterion on a toy Hamiltonian we conclude that the criterion and the 
associated functional is well-behaved.

We thank D. Vanderbilt, S. Savrasov, V. Oudovenko, and H. Jeschke for 
stimulating discussions and useful suggestions.

\appendix
\section{}
\label{sec:appen}

In this appendix we prove the assertion in Eq.~(\ref{eq:assert}).

First, if the basis is orthonormal to begin with, i.e., 
$O_{\al \be}(n-m) = \dl_{\al \be} \dl_{nm}$, it is easy to see that 
\ben
A_{\mu \si}(t) = L_{\mu  \si}(N-t) - L^{\ast}_{\si \mu}(t) = - A^{\ast}_{\si 
\mu}(t).
\een
Then, $\dl F = (4 \ep) \left| A_{\si \mu}(N-t) \right|^2 \geq 0$. 

If the basis $\{ | n \al \ra \}$ is non-orthogonal, we assume there
exists an orthonormal basis $\{ | a \tau \ra \ra \}$ (say, a Wannier basis)
to which it is related by
$
| a \tau \ra \ra = S(n, \al ; a \tau ) | n \al \ra
$
and
$
\la \la a \tau | = \la n \al | S(n, \al ; a, \tau )^{\ast}.
$
One can show that 
\ben
O^{-1}_{\al \be}(n-m) = S(n, \al ; a, \tau) S(m, \be ; a, \tau)^{\ast}.
\een
Using the above relation and Eq.~(\ref{eq:anti-h-A}) one can show that
\bea
\lefteqn{
\dl F 
}
\nonumber \\
&&
=
(4 \ep) O^{-1}_{\si \be}(m-n) A^{\ast}_{\al \be}(m) O_{\al \mu}(n-t)
A_{\mu \si}(t)
\nonumber \\
&& =
\frac{4 \ep}{N}  \sum_{a b \atop \tau \ka} \left| \sum_{t m \atop \mu \si}
S^{-1}(b, \ka ; t, \mu) A_{\mu \si}(t-m) S(m, \si ; a, \tau) \right|^2
\nonumber \\
&& \geq 0.
\eea
 

\end{document}